%% file: text.tex
\documentstyle{europhys}
\input euromacr

\begin{document}
\euro{}{}{}{}
\Date{}
\shorttitle{I. TSATSKIS: SIMPLE APPROXIMATIONS FOR ENTROPY}
\title {Simple CVM-based approximations for the configurational entropy}
\author{Igor Tsatskis
\footnote{Electronic address: it10001@cus.cam.ac.uk}
\footnote{Former name: I. V. Masanskii}}
\institute{Department of Earth Sciences, University of Cambridge,
Downing Street, Cambridge CB2 3EQ, United Kingdom}
\rec{}{}
\pacs{
\Pacs{05}{50$+$q}{Lattice theory and statistics; Ising problems}
\Pacs{61}{66Dk}{Alloys}
\Pacs{64}{60Cn}{Order-disorder transformations; statistical mechanics 
of model systems}
}
\maketitle

\begin{abstract} 
It is shown how to derive simple polynomial expressions for the 
variational configurational entropy starting from the cluster variation 
method (CVM). As an example, first six terms of the expansion of the 
entropy in powers of the nearest-neighbour (NN) short-range order (SRO) 
parameter are obtained for the NN Ising ferromagnet on the face-centered 
cubic lattice using the tetrahedron (T-CVM) approximation. Calculated 
values of the transition temperature and the NN SRO parameter at the 
transition converge rapidly to their T-CVM counterparts as order 
of the approximation increases.
\end{abstract}

The cluster variation method (CVM) proposed originally by Kikuchi~\cite{kikuchi1} 
and later reformulated, further developed and applied to a broad range of 
problems by many authors (for reviews, see~\cite{defontaine}-\cite{inden}), 
is at present a standard and well-established technique for quantitative 
calculations of finite-temperature properties of systems undergoing 
order-disorder transitions. The CVM is essentially a procedure for 
obtaining approximate analytical expressions for the variational 
configurational entropy $S$. The CVM entropy is written in terms of 
probabilities of finding various possible atomic (spin) configurations on 
lattice clusters belonging to some basic, or maximal, cluster. A 
particular CVM approximation for the entropy is therefore defined by 
the choice of the basic cluster. The second necessary ingredient, the 
internal energy $E$, is the statistical average of the Hamiltonian of 
the problem, and is a function of the same (or related) variables as the 
entropy. The resulting variational free energy $F=E-TS$ ($T$ is the absolute 
temperature) is then minimized with respect to all configurational parameters. 
This can be done using probabilities which correspond to the basic cluster.
However, they are not independent, since the sum of the probabilities of all 
possible configurations on the basic cluster is equal to unity. This condition 
is explicitly taken into account during the minimization; the obtained 
equations can then be solved by means of the natural iteration method of 
Kikuchi~\cite{kikuchi2}. Alternatively, the set of independent variables 
(spin averages) is chosen~\cite{sanchez}. The value of $F$ at the minimum 
coincides with the actual approximate free energy of the system.

Being a reliable and widely used theoretical tool, the CVM at the 
same time has certain disadvantages. One of them is that the CVM 
approximations for the entropy are fairly complicated, especially for 
comparatively large basic clusters, and the number of independent 
variational parameters in the expression for $F$ becomes progressively 
large with the increasing size of the basic cluster. As a result, the 
method is intrinsically numerical. The aim of the present study is to 
simplify the CVM thermodynamic treatment and find suitable formulae 
for the configurational entropy, thus making the variational free energy 
more tractable analytically. The general description of the proposed 
approach is followed by a particular, relatively simple example of the 
nearest-neighbour (NN) Ising ferromagnet on the face-centered cubic (FCC) 
lattice in the tetrahedron (T-CVM) approximation and by a preliminary 
test of the results obtained in this example case.

We start from the exposition of the general idea leading to the 
simplification of the CVM. In what follows the Ising language is used, 
and the variational free energy is minimized with respect to the spin 
averages which are independent configurational parameters; all relevant 
cluster probabilities can be expressed in terms of these spin averages.
The free energy thus depends on interactions $v_n$ (including the magnetic 
field) and spin averages $\xi_n$. The equilibrium values of $\xi_n$ are 
determined from the minimization conditions
\begin{equation}
\frac{\partial F}{\partial \xi_n} = 0 . \label{3}
\end{equation}
The possibility of the CVM simplification proposed here is based on the 
fact that in most CVM approximations the configurational entropy contains 
significantly more minimization parameters $\xi_n$ than the internal energy. 
The latter is the statistical average of the Hamiltonian
\begin{equation}
H = \sum_{n}{v_n a_n} , \label{4}
\end{equation}
where $a_n$ are some operators. In the theory of ordering these 
operators are products of the spin variables. The averages $\varphi_n$ 
of operators $a_n$ form a subset of the set $\{\xi_n\}$. All other spin 
averages are denoted as $\chi_n$. The internal energy then depends on 
$v_n$ and $\varphi_n$, while the entropy is a function of the spin 
averages only, but depends on both $\varphi_n$ and $\chi_n$:
\begin{equation}
E = E(\{v_n\},\{\varphi_n\}) = \sum_{n}{v_n \varphi_n} , \ \ \ 
S = S(\{\varphi_n\},\{\chi_n\}) . \label{5}
\end{equation}
Eqs.~(\ref{3}) now become
\begin{equation}
\frac{\partial S}{\partial \varphi_n} = \frac{v_n}{T} \ , \ \ \ \\
\frac{\partial S}{\partial \chi_n} = 0 \ . \label{6}
\end{equation}
It is seen that the minimization of the free energy with respect to 
the spin averages $\chi_n$ reduces to the maximization of the entropy. 
Corresponding equations are simply the relations between different 
configurational parameters, since they contain no information about 
interactions in the system, and as such can be taken into account in the 
CVM entropy from the very beginning (for a given Hamiltonian and a given 
thermodynamic phase). They can be used to find the spin averages $\chi_n$ 
as functions of the spin averages $\varphi_n$. The results are then 
substituted into the entropy, thus expressing it in terms of $\varphi_n$. 
The resulting variational free energy also depends only on the spin 
averages $\varphi_n$ conjugated to the interactions $v_n$. The number 
of remaining independent variational parameters is greatly reduced in 
comparison with the original CVM treatment; in particular, it is equal 
to the total number of distinct interactions, if all lattice sites are
equivalent ({\it i.e.}, when there are no sublattices). In the widely used 
case of pairwise interactions this approach leads to the variational 
free energy which is a function of the point and pair spin averages.

In itself the outlined approach is not very useful, since equations to be 
solved are usually rather complicated, and so would be the final expression 
for the entropy. This difficulty can be overcome by defining another set 
of parameters, the cumulants of the remaining spin averages $\varphi_n$, 
and seeking the solutions of these equations ({\it i.e.}, the expressions 
for the spin averages $\chi_n$) in the form of series expansions in powers 
of the cumulants of the pair and higher-order spin averages $\varphi_n$. 
These expansions are then substituted into the CVM approximation for the 
entropy, and the entropy itself is also expanded in powers of the same 
parameters. The cumulants of the spin averages are defined as follows:
the cumulant average of a product of $n$ spin variables is the $n$th-order 
residue when all combinations of cumulant averages of lower order are 
subtracted from the actual average of the product. In other words,
\begin{eqnarray}
& \left\langle s_i \right\rangle = \left\langle s_i \right\rangle_c , \ \ \ 
\left\langle s_i s_j \right\rangle = 
\left\langle s_i \right\rangle_c \left\langle s_j \right\rangle_c 
+ \left\langle s_i s_j \right\rangle_c , & \nonumber \\
& \left\langle s_i s_j s_k \right\rangle = 
\left\langle s_i \right\rangle_c \left\langle s_j \right\rangle_c 
\left\langle s_k \right\rangle_c + \left\langle s_i s_j \right\rangle_c 
\left\langle s_k \right\rangle_c + \left\langle s_i s_k \right\rangle_c 
\left\langle s_j \right\rangle_c + \left\langle s_j s_k \right\rangle_c 
\left\langle s_i \right\rangle_c 
+ \left\langle s_i s_j s_k \right\rangle_c , & \label{7} 
\end{eqnarray}
etc.~\cite{kubo}. Here $s_i$ is a spin variable attributed to the lattice 
site $i$, $\left\langle \right\rangle_c$ denotes the cumulant average, and 
$\left\langle \right\rangle$ is the usual statistical average. By retaining 
first several terms in the expansion for the entropy a sufficiently simple 
polynomial approximation to the (already approximate) CVM expression is 
obtained. The cumulants used as expansion parameters are almost always 
sufficiently small to justify the expansion; the irreducible pair correlation 
function, whose matrix elements are the cumulants of the pair averages and 
are proportional to the corresponding short-range order (SRO) parameters, 
can serve as an example. The coefficients of the polynomial are functions 
of the point averages. The cumulants of the point averages are these 
averages themselves, {\it i.e.}, the site magnetizations; they cannot 
be used as expansion parameters, since they are not necessarily small 
({\it e.g.}, in the ordered phase or in the case of sufficiently strong 
magnetic field above the transition).

To illustrate the proposed approach, we consider the tetrahedron version 
of the CVM for the ferromagnetic FCC Ising model with NN pair interactions. 
In the T-CVM approximation the variational configurational entropy is (see, 
{\it e.g.}, \cite{kikuchi2} or first of the reviews~\cite{defontaine})
\begin{eqnarray}
(N k_B)^{-1} S & = & - 5 \left( P^{u} \ln P^{u} + P^{d} \ln P^{d} \right) 
+ 6 \left( P^{uu} \ln P^{uu} + 2 P^{ud} \ln P^{ud} \right. \nonumber \\
& & \left. + P^{dd} \ln P^{dd} \right) - 2 \left( P^{uuuu} \ln P^{uuuu} 
+ 4 P^{uuud} \ln P^{uuud} \right. \nonumber \\
& & \left. + 6 P^{uudd} \ln P^{uudd} + 4 P^{uddd} \ln P^{uddd} 
+ P^{dddd} \ln P^{dddd} \right) . \label{8} 
\end{eqnarray}
Here superscripts ``u'' and ``d'' stand for spins up and down, $N$ is the 
number of lattice sites, and $k_{B}$ is the Boltzmann constant. The entropy 
is expressed in terms of probabilities $P$ of one-, two- and four-spin 
configurations on the basic cluster -- the NN tetrahedron. As was mentioned 
before, not all of the probabilities in eq.~(\ref{8}) are independent. The 
set of independent parameters consists of four spin averages
\begin{equation}
\xi_1 = \left\langle s_i \right\rangle = m , \ \ \ 
\xi_2 = \left\langle s_i s_j \right\rangle , \ \ \ 
\xi_3 = \left\langle s_i s_j s_k \right\rangle , \ \ \ 
\xi_4 = \left\langle s_i s_j s_k s_l \right\rangle , \label{9}
\end{equation}
where $m$ is the magnetization and all lattice sites are nearest 
neighbours. The probabilities in eq.~(\ref{8}) are statistical averages 
of the corresponding products of occupation numbers which are linearly 
related to the spin variables~\cite{defontaine}-\cite{inden}, and 
thus can be expressed in terms of averages~(\ref{9}): 
\begin{eqnarray}
& 2 P^{u} = 1 + m , \ \ \ 2 P^{d} = 1 - m , & \nonumber \\
& 4 P^{uu} = 1 + 2 m + \xi_2 , \ \ \ 
4 P^{dd} = 1 - 2 m + \xi_2 , \ \ \ 4 P^{ud} = 1 - \xi_2 , & \nonumber \\
& 16 P^{uuuu} = 1 + 4 m + 6 \xi_2 + 4 \xi_3 + \xi_4 , \ \ \ 
16 P^{dddd} = 1 - 4 m + 6 \xi_2 - 4 \xi_3 + \xi_4 , & \nonumber \\
& 16 P^{uuud} = 1 + 2 m - 2 \xi_3 - \xi_4 , \ 
16 P^{uddd} = 1 - 2 m + 2 \xi_3 - \xi_4 , \
16 P^{uudd} = 1 - \xi_2 + \xi_4 . \ \ \ & \label{25}
\end{eqnarray}
The Hamiltonian of the Ising model is
\begin{equation}
H = - \frac{1}{2} \sum_{ij}{J_{ij} s_i s_j} - h \sum_{i}{s_i} \label{10}
\end{equation}
($J_{ij}$ is the exchange integral equal to $J>0$ for NN sites and to 
zero otherwise, and $h$ is the magnetic field), so that only first two 
averages~(\ref{9}) enter the expression for the internal energy, 
\begin{equation}
N^{-1} E = - \frac{1}{2} Z J \xi_2 - h \xi_1 , \label{11}
\end{equation}
where $Z$ is the NN coordination number. As a result, the set $\{\xi_n\}$ 
contains four averages, while its subset $\{\varphi_n\}$ consists of two 
variables $\xi_1$ and $\xi_2$. The minimization of the free energy with 
respect to $\xi_3$ and $\xi_4$ reduces to the maximization of the entropy 
with respect to these two variables and leads to the following equations:
\begin{equation}
P^{uuuu} \left( P^{uddd} \right)^2 = 
P^{dddd} \left( P^{uuud} \right)^2 , \ \ \ 
P^{uuuu} P^{dddd} \left( P^{uudd} \right)^6 = 
\left( P^{uuud} P^{uddd} \right)^4 . \label{12}
\end{equation}
From eqs.~(\ref{12}) the averages $\xi_3$ and $\xi_4$ can be found as 
functions of the averages $\xi_1$ and $\xi_2$. According to the proposed 
general scheme, $\xi_3$ and $\xi_4$ should be sought as series expansions 
in powers of the cumulant of the pair average $\xi_2$. This cumulant is 
the NN matrix element of the irreducible pair correlation function. However, 
a quantity proportional to it, the NN SRO parameter $\alpha$, is more often 
used. The relation between $\xi_2$ and $\alpha$ is 
\begin{equation}
\xi_2 = m^2 + (1-m^2) \alpha . \label{13}
\end{equation}
By iteratively solving eqs.~(\ref{12}) first six terms of the series expansions 
in powers of $\alpha$ for the spin averages $\xi_3$ and $\xi_4$ were found,
\begin{eqnarray}
& \xi_3 = a_0 + a_1 \alpha + a_2 \alpha^2 + a_3 \alpha^3 
          + a_4 \alpha^4 + a_5 \alpha^5 + a_6 \alpha^6 
          + O(\alpha^{7}) , & \label{14a} \\
& \xi_4 = b_0 + b_1 \alpha + b_2 \alpha^2 + b_3 \alpha^3 
          + b_4 \alpha^4 + b_5 \alpha^5 + b_6 \alpha^6 
          + O(\alpha^{7}) , & \label{14b}
\end{eqnarray}
where the coefficients $a_n$ and $b_n$ are
\begin{eqnarray}
& a_0 = m^3 , \ \ \ a_1 = 3 m \left( 1 - m^2 \right) , \ \ \ 
a_2 = -6 m \left( 1 - m^2 \right) , & \nonumber \\
& a_3 = 2 m \left( 5 - 9 m^2 \right) , \ \ \ 
a_4 = -6 m \left( 1 - 9 m^2 \right) , \ \ \ 
a_5 = -54 m \left( 1 - m^2 \right) , & \nonumber \\
& a_6 = 2 m \left( 1 - m^2 \right)^{-2} 
\left( 185 - 1491 m^2 + 3051 m^4 - 1809 m^6 \right) , & \nonumber \\
& b_0 = m^4 , \ \ \ b_1 = 6 m^2 \left( 1 - m^2 \right) , \ \ \ 
b_2 = 3 \left( 1 - m^2 \right) \left( 1 - 9 m^2 \right) , & \nonumber \\
& b_3 = -8 \left( 1 - 15 m^2 + 18 m^4 \right) , \ \ \ 
b_4 = 24 \left( 1 - 19 m^2 + 36 m^4 \right) , & \nonumber \\ 
& b_5 = -24 \left( 1 - m^2 \right)^{-1} 
\left( 3 - 70 m^2 + 267 m^4 - 216 m^6 \right) , & \nonumber \\ 
& b_6 = 8 \left( 1 - m^2 \right)^{-2} 
\left( 26 - 691 m^2 + 4089 m^4 - 6993 m^6 + 3537 m^8 \right) . & \label{16}
\end{eqnarray}
Note that the expressions for the first four coefficients $a_n$ and $b_n$ 
coincide with the exact results obtained in the case of arbitrary pairwise 
interactions on the FCC lattice~\cite{masanskii}. The tetrahedron CVM therefore 
correctly reproduces exact expansions for the averages $\xi_3$ and $\xi_4$ 
up to at least third order in $\alpha$; exact formulae for the higher-order 
coefficients are at present unavailable. Substitution of the expansions~(\ref{14a}) 
and (\ref{14b}) into the original T-CVM formula~(\ref{8}) and subsequent 
expansion of the entropy in powers of $\alpha$ gives 
\begin{eqnarray}
& S = S_0 + S_1 , & \label{17a} \\
& (N k_B)^{-1} S_0 = 
- \left( \frac{1+m}{2} \ln \frac{1+m}{2} + 
\frac{1-m}{2} \ln \frac{1-m}{2} \right) , & \label{17b} \\
& (N k_B)^{-1} S_1 = c_2 \alpha^2 + c_3 \alpha^3 + c_4 \alpha^4 
+ c_5 \alpha^5 + c_6 \alpha^6 + O(\alpha^{7}) , & \label{17c} \\
& c_2 = - 3 , \ \ 
c_3 = 4 \left( 1 - m^2 \right)^{-1} \left( 2 - m^2 \right) , \ \ 
c_4 = - 0.5 \left( 1 - m^2 \right)^{-2} 
\left( 37 - 66 m^2 + 45 m^4 \right) , & \nonumber \\
& c_5 = 4.8 \left( 1 - m^2 \right)^{-3} 
\left( 10 - 39 m^2 + 52 m^4 - 19 m^6 \right) , & \nonumber \\
& c_6 = - 0.2 \left( 1 - m^2 \right)^{-4} \left( 641 - 3900 m^2 
+ 7870 m^4 - 6220 m^6 + 1865 m^8 \right) . & \label{18}
\end{eqnarray}
As expected, the entropy is a sum of two terms, the mean-field 
(Bragg-Williams) approximation (MFA) contribution $S_0$ and the 
correlation entropy $S_1$. The mean-field part of the entropy is the 
zero-order term, the linear term vanishes, since the entropy is maximal 
in the case of uncorrelated spins, and the expansion of the correlation 
entropy starts from the quadratic term. Now the variational free 
energy is a very simple function of only two parameters $m$ and $\alpha$ 
(instead of four variables~(\ref{9})). The equilibrium values of these 
two parameters are the solutions of two minimization equations, 
\begin{equation}
\frac{\partial F}{\partial m} = 0 \ , \ \ \ 
\frac{\partial F}{\partial \alpha} = 0 \ . \label{19}
\end{equation}
The obtained result~(\ref{17c}) for the correlation entropy has 
especially simple form in the disordered phase with zero magnetic 
field. In this case first of eqs.~(\ref{19}) leads to the vanishing 
magnetization, while second gives $\alpha$ as a function of a single 
parameter -- the normalized temperature $t = k_B T / ZJ$; the 
expansion~(\ref{17c}) has numeric coefficients:
\begin{equation}
(N k_B)^{-1} S_1 = - 3 \alpha^2 + 8 \alpha^3 - 18.5 \alpha^4 + 48 \alpha^5 
- 128.2 \alpha^6 + O(\alpha^{7}) \ . \label{21}
\end{equation}

In the polynomial approximations for the entropy which can be obtained 
from the expansion~(\ref{17c}) by retaining finite number of terms, 
the highest-order term must always be the even-order one. Otherwise, 
from the positiveness of the third- and fifth-order coefficients in the 
expansion~(\ref{21}) it follows that in the disordered phase the approximate 
entropy does not decrease monotonically with increasing positive $\alpha$, 
but has a minimum at some $\alpha$ value and begins to increase afterwards. 
As a result, the solution of the second of eqs.~(\ref{19}) disappears as 
temperature decreases, while at higher temperatures this equation has two 
solutions instead of one. On the other hand, all calculated even-order 
coefficients are negative, and no such difficulties occur. As a preliminary 
test of the expansion~(\ref{17c}) for the correlation entropy and related 
approximations for the variational free energy obtained in the present study, 
we calculate the normalized transition temperature $t_c$ and the NN SRO parameter 
at the transition point $\alpha_c$. The transition temperature is defined by 
the condition $\partial^2 F / \partial m^2 = 0$, together with the minimization 
equations~(\ref{19}). Values of $t_c$ and $\alpha_c$ calculated in the three 
available (second-, fourth- and sixth-order) approximations are given in 
Table~\ref{t1}. They clearly converge to the original T-CVM values, from above 
for $t_c$ and from below for $\alpha_c$; for instance, the overestimation of 
$t_c$ changes from 20\% in the MFA to 9\% in the quadratic approximation, and 
then to 3\% in the highest (sixth-order) approximation.

\begin{table}
\caption{Normalised transition temperature $t_c = k_B T_c / Z J$ and 
the NN SRO parameter at this temperature $\alpha_c = \alpha (t_c)$ 
in the three approximations for the variational free energy obtained 
from expansion~(\ref{17c}) for the correlation entropy. The MFA and 
T-CVM values, as well as corresponding ratios, are also shown for the 
purpose of comparison. Results for the transition temperature can be 
further compared with the best known value $t_c = 0.81638$~\cite{defontaine}.}
\label{t1}
\begin{center}
\begin{tabular}{ccccc}
\hline
Approximation & $t_c$ & $t_c / t_c^{T-CVM}$ & $\alpha_c$ & 
$\alpha_c / \alpha_c^{T-CVM}$ \\
\hline
MFA (0th order) &    1    & 1.197 &    0    &   0   \\
2nd order       & 0.90825 & 1.087 & 0.09175 & 0.459 \\
4th order       & 0.87703 & 1.050 & 0.13930 & 0.697 \\
6th order       & 0.86205 & 1.032 & 0.16304 & 0.815 \\
T-CVM           & 0.83545 &   1   & 0.20000 &   1   \\
\hline
\end{tabular}
\end{center}
\end{table}

In summary, we have proposed the method of obtaining simple polynomial 
approximations for the variational configurational entropy which is 
based on the widely used CVM. This method utilizes the fact that 
the CVM entropy usually depends on much greater number of variational 
parameters than the internal energy. The minimization of the free energy 
with respect to variables which do not enter the expression for the 
internal energy is thus equivalent to the maximization of the entropy.
The corresponding equations do not contain any information about the 
Hamiltonian of the system and play the role of constraints relating different 
variational parameters. They are used for the elimination of the redundant 
variables. The number of the remaining independent variational parameters 
in the free energy is almost always significantly reduced in comparison with 
the CVM. In the case of the absence of sublattices in the system it is equal 
to the total number of different interactions in the Hamiltonian (the magnetic 
field included). In practice this programme is implemented by using the 
expansion in powers of the cumulants of the remaining pair and higher-order 
spin averages: (i) the variables to be eliminated are found from the 
constraint equations as the corresponding series, (ii) the results are 
substituted into the initial CVM formula for the entropy, (iii) the obtained 
expression for the entropy is also expanded into the same series and (iv) 
finite number of terms in this expansion is retained. The resulting 
approximate free energy is a polynomial in the cumulants of the pair and 
higher-order spin averages conjugated to the interactions, {\it i.e.}, 
present in the expression for the internal energy. The coefficients of the 
polynomial are functions of the (sub)lattice magnetization(s). The final 
expression for the free energy always has the same functional form, 
regardless of the particular CVM approximation used initially for the 
configurational entropy; only the coefficients of the polynomial change. 
The evolution of these coefficients with the version of the CVM is a matter 
of considerable interest and will be the subject of a separate study. As a 
particular example, the ferromagnetic FCC Ising model with NN pair interactions 
in the tetrahedron approximation has been considered. First six terms of the 
expansion of the configurational entropy in powers of the NN SRO parameter 
have been obtained. Calculated values of the transition temperature and the 
NN SRO parameter at this temperature rapidly converge to the corresponding 
original T-CVM results with the increasing order of the approximation. The 
detailed study of the applicability and accuracy of the obtained approximations 
for the free energy is in progress and will be published elsewhere.

\stars

\end{document}

%% file: euromacr.tex
\def\And{{\rm and\ }}

\def\stars{\bigskip\centerline{***}\medskip}

\newif\ifboo \boofalse

\def\Review#1{\boofalse{\it #1},}
\def\Name#1{{\sc #1},}
\def\Vol#1{\ifboo Vol. {\bf #1}\else{\bf #1}\fi}
\def\Year#1{\ifboo #1\else(#1)\fi}
\def\Book#1{\bootrue{\it #1},}
\def\Page#1{\ifboo {\rm p. #1}\else{\rm #1}\fi}

%% file: text.bbl
\vskip-12pt
\begin{thebibliography}{99}

\bibitem{kikuchi1}
\Name{Kikuchi R.} \Review{Phys. Rev.} \Vol{79} \Year{1950} \Page{718}; 
\Vol{81} \Year{1951} \Page{988}.

\bibitem{defontaine}
\Name{de Fontaine D.} \Review{Solid State Physics} \Vol{34} \Year{1979} \Page{73};
\Vol{47} \Year{1994} \Page{33}.

\bibitem{ducastelle} 
\Name{Ducastelle F.} \Book{Order and Phase Stability in Alloys} 
(North-Holland, Amsterdam) \Year{1991}.

\bibitem{inden} 
\Name{Inden G. \And Pitsch W.} in \Book{Phase Transformations in Materials} 
edited by \Name{P. Haasen} (VCH Press, New York) \Year{1991} p\Page{497-552}.

\bibitem{kikuchi2}
\Name{Kikuchi R.} \Review{J. Chem. Phys.} \Vol{60} \Year{1974} \Page{1071}. 

\bibitem{sanchez} 
\Name{Sanchez J.M. \And de Fontaine D.} \Review{Phys. Rev. B} 
\Vol{17} \Year{1978} \Page{2926}.

\bibitem{kubo}
\Name{Kubo R.} \Review{J. Phys. Soc. Japan} \Vol{17} \Year{1962} \Page{1100}.

\bibitem{masanskii}
\Name{Masanskii I.V. \And Tokar V.I.} \Review{J. Phys. I France} 
\Vol{2} \Year{1992} \Page{1559}.

\end{thebibliography}
